\documentclass[journal=jctc,manuscript=article]{achemso}
\usepackage[version=3]{mhchem} % Formula subscripts using \ce{}
\usepackage{amssymb}
\usepackage{amsmath}
\usepackage{hyperref}
\usepackage[dvipsnames]{xcolor}
\usepackage[capitalize]{cleveref}
\usepackage{algorithm}
\usepackage{algorithmic}
\SectionNumbersOn

\author{Maaike M. Galama}
\affiliation{Department of Mathematics and Computer Science, Freie Universit\"{a}t Berlin, Arnimallee 12, 14195 Berlin, Germany}
\author{Hao Wu}
\email{hwu81@sjtu.edu.cn}
\affiliation{School of Mathematical Sciences, Institute of Natural Sciences, and MOE-LSC, Shanghai Jiao Tong University, 200240 Shanghai, China}
\alsoaffiliation{School of Mathematical Sciences, Tongji University, 200092 Shanghai, China}
\author{Andreas Kr\"{a}mer}
\email{andreas.kraemer@fu-berlin.de}
\affiliation{Department of Mathematics and Computer Science, Freie Universit\"{a}t Berlin, Arnimallee 12, 14195 Berlin, Germany}
\author{Mohsen Sadeghi}
\affiliation{Department of Mathematics and Computer Science, Freie Universit\"{a}t Berlin, Arnimallee 12, 14195 Berlin, Germany}
\author{Frank Noé}
\email{frank.noe@fu-berlin.de}
\affiliation{Department of Mathematics and Computer Science, Freie Universit\"{a}t Berlin, Arnimallee 12, 14195 Berlin, Germany}
\alsoaffiliation{Department of Physics, Freie Universit\"{a}t Berlin, Arnimallee 14, 14195 Berlin, Germany}
\alsoaffiliation{Department of Chemistry, Rice University, Houston, TX, USA}
\alsoaffiliation{Microsoft Research, Cambridge, UK}

\title[]{Stochastic approximation to MBAR and TRAM: batch-wise free energy estimation}

\abbreviations{}

\begin{document}

\renewcommand{\thesection}{{}}
\renewcommand{\thesubsection}{}

\begin{abstract}
The dynamics of molecules are governed by rare event transitions between long-lived (metastable) states. To explore these transitions efficiently, many enhanced sampling protocols have been introduced that involve using simulations with biases or changed temperatures. Two established statistically optimal estimators for obtaining unbiased equilibrium properties from such simulations are the multistate Bennett Acceptance Ratio (MBAR) and the transition-based reweighting analysis method (TRAM). Both MBAR and TRAM are solved iteratively and can suffer from long convergence times. Here we introduce stochastic approximators (SA) for both estimators, resulting in SAMBAR and SATRAM, which are shown to converge faster than their deterministic counterparts, without significant accuracy loss. Both methods are demonstrated on different molecular systems.
\end{abstract}

\begin{tocentry}
   \includegraphics[width=8.25cm, height=4.45cm]{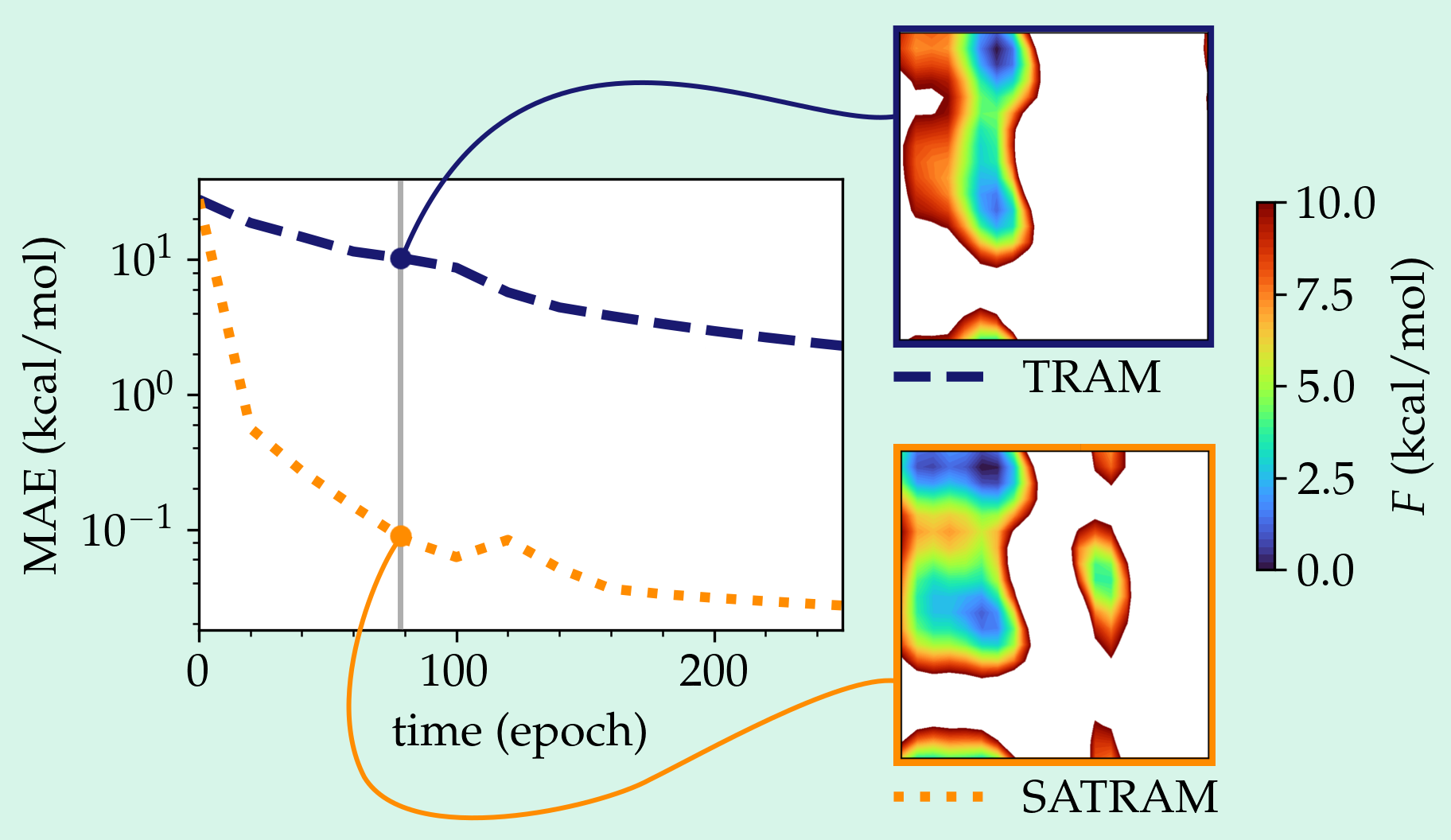}
\end{tocentry}

\section*{Introduction}

Computer simulations are a vital tool in biochemistry and biophysics. They complement and even replace physical experiments for investigating important biomolecular processes such as protein-ligand binding \cite{shukla2018identification,hempel2021synergistic,novack2022binding}, protein folding\cite{piana2014assessing,li2018finding,sugita1999replica}, and membrane permeation \cite{de2001water,venable2019molecular,gupta2020permeation,sugita2021large}. However, many of these processes involve complex transitions that are so slow with respect to the simulation time scale, that they might not be observed even once within a simulation time span \cite{voelz2010molecular, voelz2012slow}.

Enhanced sampling methods modify the Hamiltonian of a system in such a way that one is able to sample the thermodynamics of slow processes more efficiently, for example by adding biasing potentials (e.g. umbrella sampling\cite{torrie1977nonphysical,bartels1999adaptive}, accelerated molecular dynamics\cite{hamelberg2004accelerated,miao2015gaussian}), changing state parameters such as the temperature (e.g. replica exchange\cite{swendsen1986replica,sugita1999replica})\cite{hansmann1997parallel,marinari1992simulated}, or metadynamics \cite{grubmuller1995predicting,barducci2011metadynamics,voter1997hyperdynamics}. 

In order to un-bias the samples from such simulations, reweighting estimators combine data sampled at different thermodynamic states to compute observables in the state of interest. Important estimators include the weighted histogram analysis method (WHAM) \cite{ferrenberg1989optimized,kumar1992weighted,souaille2001extension} and bin-less WHAM, also known as multistate Bennett acceptance ratio (MBAR) \cite{shirts2008statistically,kong2003theory,tan2012theory,zhang2015stochastic,bartels2000analyzing}.
When combining the estimation of different thermodynamic states with Markov state models, one obtains a family of estimators that include the discrete transition-based analysis method (dTRAM) \cite{wu2014statistically}, the dynamic weighted historam analysis method (DHAM) \cite{rosta2015free} and the transition-based analysis method (TRAM)\cite{wu2016multiensemble}. In this paper we concentrate on MBAR and TRAM.

MBAR is a statistically optimal method to combine data from multiple thermodynamic states \cite{kong2003theory,bartels2000analyzing,shirts2008statistically}. It is a generalization of the Bennett acceptance ratio method \cite{bennett1976efficient} to an arbitrary number of states, has been shown to minimize the estimator variance \cite{shirts2008statistically} and works under the assumption that all states were sampled in equilibrium. The latter condition can be difficult to achieve for molecular systems with rare events.

TRAM \cite{wu2016multiensemble} combines reweighting from different thermodynamic states with Markov State Model (MSM) \cite{schutte1999direct,swope2004describing,singhal2005error,sriraman2005coarse,noe2007hierarchical,chodera2007automatic,prinz2011markov} theory to enable free energy estimation from trajectories that are too short to have produced samples that reflect the equilibrium distribution. The trajectory samples only need to be in a local equilibrium, but not in global equilibrium, allowing for more arbitrary choices of initial conditions and use of short trajectories. Moreover, TRAM extracts kinetic information in the form of an MSM in addition to thermodynamic information in the form of free energy estimates.

Both MBAR and TRAM were originally formulated in the form of self-consistent equations that can be solved iteratively \cite{shirts2008statistically,wu2016multiensemble}. Depending on the system and data under consideration, these self-consistent algorithms can take hours or even days to converge \cite{zhang2015stochastic,ding2019fast,tan2016locally}. To accelerate convergence, MBAR has been adapted and optimized in various ways since its inception \cite{zhang2015stochastic,zhang2017stratified,ding2019fast,jia2021free,ferguson2017bayeswham,tan2016locally}, but for TRAM the self-consistent algorithm remains the only available solver as yet.

In this work, we develop SATRAM, a stochastic approximation for TRAM, which converges faster to a given accuracy threshold than TRAM, decreasing estimation time up to an order of magnitude. By extension, we also develop SAMBAR, a stochastic approximation for MBAR, which is a special case of SATRAM in the same way that MBAR is a special case of TRAM: they are identical when all data comes from a single Markov state.

\section*{Theory}
Our system of interest is a reference ensemble governed by a dimensionless potential function $u(x)$ ($x$ being a configuration state), which has units of thermal energy, $\beta=(k_B T)^{-1}$, $T$ being the temperature. In the canonical ensemble, $u(x)$ is equal to $\beta U(x)$, where $U$ is the potential energy function. 
The equilibrium distribution of the system is given by the Boltzmann distribution
\begin{equation}
    \mu(x) = e^{f-u(x)},
\end{equation}
where $f$, the free energy, acts as a constant that normalizes $\mu(x)$, and is equal to the negative log of the partition function.

Now we consider the case where we have collected samples from $K$ simulations that were performed at different thermodynamic states. Each thermodynamic state can be related to our unbiased (reference) ensemble as $u^k(x) = u(x) + b^k(x)$, $k \in \{1, ... K\}$ being the thermodynamic state index. The term $b^k(x)$ can result from various thermodynamic changes such as an added umbrella potential, alchemical potentials or a change in temperature, but will from here on simply be referred to as the \textit{bias} energy. The equilibrium distribution for each state can be written as a reweighting of the unbiased distribution as
\begin{equation}
    \label{eq:unbiased_probability}
    \mu^k(x) = e^{f^k - b^k(x)}\mu(x),
\end{equation}
where the $f^k$ are the free energies of the different ensembles, and again function as normalizing factors. Our goal is to estimate the $f^k$ so that we can compare the free energies of the biased states, and/or for the purpose of being able to reweight our samples back to the unbiased distribution using Eq. (\ref{eq:unbiased_probability}).

\subsection*{MBAR}
MBAR is equivalent to binless WHAM \cite{shirts2008statistically,kong2003theory,bartels2000analyzing} and can be derived as a generalization of the Bennett Acceptance Ratio (BAR) \cite{bennett1976efficient} to multiple states. MBAR thus reduces to BAR when only two states are considered.

The self-consistent MBAR equations are given by
\begin{equation}
f^k = -\ln\sum_{x\in X} \frac{\exp[-b^k(x)]}{\sum_{l=1}^K N^l \;\exp[f^l - b^l(x)]},
\label{eq:MBAR_self_consistent}
\end{equation}
where $N^k$ are the number of samples taken at thermodynamic state $k$, and $X$ is the set of all samples of size $|X| = N = \sum_k N^k$. 
These equations can be solved by iterating until self-consistency is achieved. Since the equations depend on all samples, the computational cost for a single iteration scales as $O(KN)$, i.e., proportional to the total number of samples $N$ and total number of thermodynamic states $K$. The same scaling behavior also applies to numerical algorithms that are based on maximum-likelihood formulations of MBAR \cite{ding2019fast}.

\subsection*{TRAM}
The transition-based reweighting analysis method, TRAM \cite{wu2016multiensemble}, combines MSMs and MBAR and generalizes both. TRAM assumes that the trajectories sampled in each thermodynamic states can be approximated by a Markov model, thereby relaxing MBAR's assumption that the data samples from global equilibrium and allowing for shorter trajectories. This is modeled by the assumption that within each Markov state, data is distributed according to a local equilibrium distribution $\mu_i^k(x)$, where $i$ is the Markov state index and $k$ the thermodynamic state index. The local equilibrium distribution can be reweighted to the global equilibrium distribution, $\mu(x)$ by the relation
\begin{equation}
    \mu_i^k(x) =  \mu(x) e^{f_i^k -b^k(x)},
\end{equation}
where $f_i^k$ is the free energy of configuration state $i$ in thermodynamic state $k$, and $b^k(x)$ is the bias energy of sample $x$ evaluated at thermodynamic state $k$.

The TRAM likelihood thus consists of two parts. It combines the MSM likelihood of sampling a specific sequence of discrete states ($L_{\mathrm{MSM}}$), with the local equilibrium likelihood of sampling the continuous configurations within those states ($L_{\mathrm{LEQ}})$. The total likelihood is the product over all thermodynamic states, and given by \cite{wu2016multiensemble}
\begin{equation}
L_{\mathrm{TRAM}} = \prod_{k=1}^K \underbrace{\left( \prod_{i,j}(p_{ij}^k)^{c_{ij}^k} \right)}_{L^k_{\mathrm{MSM}}} \underbrace{\left( \prod_{i=1}^m \prod_{x \in X_i^k} \mu(x) e^{f_i^k -b^k(x)} \right)}_{L^k_{\mathrm{LEQ}}},
\end{equation}
where $K$ is the number of thermodynamic states, and $m$ the number of configuration states. The MSM term in the likelihood is the product of the $p_{ij}^k$, the probability of transitioning from configuration state $i$ to configuration state $j$ in thermodynamic state $k$, to the power of $c_{ij}^k$, the observed number of transitions, for all pairs of configuration states $i$ and $j$. The local-equilibrium term is formed by reweighting the aforementioned local equilibrium distribution $\mu_i^k(x)$, where $X_i^k$ is the set of samples sampled from thermodynamic state $k$ that fall into Markov state $i$. 

The TRAM solution is obtained by maximizing this likelihood with appropriate normalization constraints and with the detailed balance constraints $e^{f_i^k} p^k_{ij}=e^{f_j^k} p^k_{ji}$.
\citet{wu2016multiensemble} show that the maximum-likelihood estimate for $f_i^k$ and $v_i^k$ satisfies the following equations:
\begin{align}
\sum_j \frac{c_{ij}^k + c_{ji}^k}{\mathrm{exp}[f_j^k - f_i^k] v_j^k + v_i^k} = 1, \quad \text{for all } \;i,\; k, \label{eq:TRAM_1}\\
\sum_{x \in X_i} \frac{\mathrm{exp}[f_i^k-b^k(x)]}{\sum_l R_i^l \mathrm{exp}[f_i^l - b^l(x)]} = 1, \quad \text{for all }\; i, \;k.
\label{eq:TRAM_2}
\end{align}
The $v_i^k$ are a matrix of Lagrange multipliers that arise by reducing the TRAM problem from a constrained problem to an unconstrained problem. The $R_i^k$ are effective state counts given by
\begin{equation}
R_i^k = \sum_j \frac{(c_{ij}^k + c_{ji}^k)v_j^k}{v_j^k+\mathrm{exp}[f_i^k -f_j^k]v_i^k}  + N_i^k - \sum_j c_{ji}^k,
\label{eq:reduced_sample_counts}
\end{equation}

The computational bottleneck in these equations are the equations \cref{eq:TRAM_2} which, like the MBAR equations, depend on all samples and thereby also have a computational complexity of order $O(KN)$.
TRAM is equivalent to MBAR when all samples fall into a single Markov state.

\subsection*{Related work}
Before introducing the stochastic methods SAMBAR and SATRAM, we note that there are various existing implementations and optimizations available for MBAR. Perhaps the most well-known implementation is pymbar \cite{shirts2008statistically}, which combines self-consistent iteration of the MBAR equations with treating MBAR as an optimization problem for maximum precision. The FastMBAR \cite{ding2019fast} package implements MBAR as an optimization problem only and runs faster owing to its GPU-accelerated Quasi-Newton solver but does not boast the extensive functionality of pymbar, such as the computation of state overlap and observables. Both make use of scipy's \cite{2020SciPy-NMeth} optimization algorithms.

An existing stochastic solver of the MBAR equations is RE-SWHAM \cite{zhang2015stochastic}, which uses a post-hoc replica-exchange algorithm that re-samples the data obtained from multiple simulations. Stratified UWHAM and its stochastic counterpart stratified RE-SWHAM build on this by being able to handle systems out of equilibrium \cite{zhang2017stratified}. This is done by expanding the set of thermodynamic states by splitting data that was sampled out of equilibrium into multiple thermodynamic states that are (artificially) in equilibrium, after which MBAR can be applied. 

Local WHAM \cite{tan2017optimally} and its stochastic counterpart resample data using generalized serial tempering, and approximate the free energies by only considering neighbouring states, thereby reducing computational complexity. Similarly, divide-and-conquer MBAR (DC-MBAR) \cite{jia2021free} only calculates the free energy differences between states which have sufficient overlap, which reduces the computational complexity from $O(KN)$ to $O(N).$

For TRAM, the original implementation is available as part of the PyEMMA software package \cite{scherer_pyemma_2015}. Additionally, the Deeptime \cite{hoffmann2021deeptime} functionality has recently been extended with a parallelized TRAM implementation. The authors are not aware of any stochastic implementation, or any other optimization attempt with respect to TRAM.

\subsection*{Stochastic approximation}
Both MBAR and TRAM are formulated as nonlinear equation systems, Eqs. \eqref{eq:MBAR_self_consistent} and \eqref{eq:TRAM_1}-\eqref{eq:TRAM_2}, respectively, which we here solve by stochastic approximation.
To introduce this formalism, let $M(x)$ be a function and $a$ a constant such that the equation
\begin{equation}
M(x) = a
\end{equation}
has a unique solution at $x = \hat{x}$. \citet{robbins1951stochastic} showed that if we can observe a random variable, $N(x)$, for which $\mathbb{E}[N(x)] = M(x)$, we can approximate $\hat{x}$ by iteratively solving
\begin{equation}
    \label{eq:SA}
    x_{n+1} = x_n - \eta_n(N(x_n) - a),
\end{equation}
where $n$ is the iteration number, and $\eta$ the learning rate which controls the size of the parameter update. It can be shown that, under mild assumptions, $x$ will almost surely converge to the optimal value $\hat{x}.$

In machine learning practice, the random variable $N(x)$ is usually the gradient of an objective function evaluated over a small randomly chosen subset of the entire dataset, called a \textit{batch}. Evaluating the objective function with respect to the batch is computationally cheaper than evaluating it over the entire dataset, so that stochastic iterations are faster than their deterministic counterparts.

Stochastic approximation was previously used by \citet{tan2017optimally} to derive the locally weighted histogram analysis method, a stochastic approximation to WHAM.

\subsection*{SAMBAR}
We apply the stochastic approximation from Eq. \eqref{eq:SA} to MBAR, see supplementary information for the full derivation.
The resulting stochastic method iteratively draws a random batch of samples of size $|B|$ and performs the update
\begin{equation}
f^k = f^k - \eta \left( \frac{1}{|B|} \sum_{x\in B} \frac{N \; \exp[f^k -b^k(x)]}{\sum_{l=1}^S N^l \;\exp[f^l - b^l(x)]}\right),
\end{equation}
% \begin{equation}
% f^k = f^k - \eta \left( \frac{1}{|B|} \sum_{x\in B} \frac{N^k \exp[f^k -b^k(x)]}{\sum_{l=1}^S N^l \;\exp[f^l - b^l(x)]} - \frac{N^k}{N} \right),
% \end{equation}
where $\eta$ is the learning rate. The batchwise algorithm reduces the computational complexity for one parameter update to $O(K)$. The expression between the brackets in fact equals the gradient of the MBAR likelihood of observing the batch of samples, i.e. equivalent to performing batch-wise gradient descent. When $|B|=N$, this is equivalent to solving MBAR as a convex optimization problem as shown in \citet{shirts2008statistically,ding2019fast}.

\subsection*{SATRAM}
Analogously, we also derive stochastic approximation for the TRAM equations, see supplementary information. SATRAM updates the free energies $f_i^k$ and Lagrange multipliers $v_i^k$ as  
\begin{align}
    f_i^k &:= f_i^k - \eta \left(\frac{1}{|B|} \sum_{x \in B} 1_{x\in S_i}  \frac{\mathrm{exp}[f_i^k-b^k(x)]}{\sum_l R_i^l\; \mathrm{exp}[f_i^l - b^l(x)]} \right), \label{eq:SATRAM_1}\\
    v_i^k &:= (1-\eta)\cdot v_i^k + \eta  \cdot \frac{1}{N} \sum_j \frac{(c_{ij}^k + c_{ji}^k)v_i^k}{\mathrm{exp}[f_j^k - f_i^k] v_j^k + v_i^k} \label{eq:SATRAM_2},
\end{align}

where the $R_i^k$ are computed as 
\begin{equation}
    R_i^k = \frac{1}{N} \left(\sum_j \frac{(c_{ij}^k + c_{ji}^k)v_j^k}{v_j^k+\mathrm{exp}[f_i^k -f_j^k]v_i^k} + N_i^k - \sum_j c_{ji}^k \right).
\end{equation}
The computational complexity of the update of the $f_i^k$ has been reduced to $O(K)$. 

To avoid a large drift in the energies, a normalization step is executed after each free energy update, which consists of shifting all $f_i^k$ by a constant so that the minimum of the free energies equals zero.

When all samples are binned into a single Markov state all indices $i$ and $j$ refer to the single state and most factors in  $R_i^k = R^k$ cancel out so that $R_i^k$ reduces to $\frac{N^k}{N}$. The SATRAM equations thus reduce to SAMBAR in the same way that TRAM reduces to MBAR when considering only one Markov state.

\subsection*{Algorithmic details}

\paragraph{Initialization} 
The runtime until convergence of (SA)TRAM can be improved by starting estimation from the mean of the bias energies, i.e. $f_i^k = \langle b^k(x) \rangle, \; \forall k, i$ (see also the implementation notes in the supplementary information).

\paragraph{Batch size scheduling} The hyperparameters in the free energy update in Eq. \eqref{eq:SATRAM_1} are the learning rate $\eta$ and the batch size $b=|B|$. Setting the learning rate to $\eta=\sqrt{b/N}$ and increasing the batch size over time was found to be more robust than varying the learning rate (see supplementary information). The initial batch size $b_0$ is doubled every $p$ epochs. After each doubling of the batch size, the learning rate is again set to $\eta=\sqrt{b/N}$. When the batch size reaches the total dataset size, SATRAM reverts to a deterministic implementation. The learning rate may be decreased over time to achieve convergence to arbitrary accuracy, though in all our experiments decreasing the learning rate was not necessary to converge to well below chemical accuracy.

\paragraph{Finalization}
For all configuration states $i$ that were \textit{not} sampled in thermodynamic state $k$, $R_i^k=0$. TRAM estimates a finite free energy for these empty states based on the estimate in states where $R_i^l > 0$, $l \neq k$, but SATRAM does not, since the division by $R_i^k$ in \cref{eq:SATRAM_1} would cause numerical errors for empty states. Although this does not affect computed observables, it does affect the estimated $f^k$, where $f^k =-\mathrm{ln}\sum_i \mathrm{exp}[-f_i^k]$. To estimate the free energy of these empty states,
\begin{equation}
f_i^k = -\mathrm{ln} \sum_{x \in X_i} \frac{\mathrm{exp}[-b^k(x)]}{\sum_l R_i^l \mathrm{exp}[f_i^l - b^l(x)]}    \label{eq:SATRAM_finalize}
\end{equation}
is computed once as a finalization step after SATRAM has converged. This equation is equal to the second step of the TRAM algorithm, see Eq. (19) in \citet{wu2016multiensemble}.
 
The full SATRAM algorithm is summarized in \cref{alg:SATRAM}. 
\begin{algorithm}
\caption{The SATRAM Algorithm}\label{alg:SATRAM}
\begin{algorithmic}
\STATE \textbf{parameters:} initial batch size $b$, doubling interval $p$, initialized $f_i^k$ and $v_i^k$
\WHILE{($n <$ max iterations) and (convergence criterion not reached)}
    \STATE $B \gets$ random batch of samples, $|B|=b$
    \STATE update $f_i^k$ using \cref{eq:SATRAM_1} with $B$ and $\eta = \sqrt{b/N}$
    \STATE update $v_i^k$ using \cref{eq:SATRAM_2}
    \IF{$n$ mod $p == 0$}
    \STATE $b \gets 2\cdot b$
    \ENDIF
    \STATE $n \gets n+1$
\ENDWHILE
\STATE finalize $f_i^k$ using \cref{eq:SATRAM_finalize}
\end{algorithmic}
\end{algorithm}

\section*{Results and discussion}
We compare the performance of SATRAM and TRAM by applying them to three different sets of simulation data. The first two dataset both contain alanine dipeptide simulation data, generated by simulations at multiple temperatures and by umbrella sampling simulations respectively. The third dataset was generated by performing umbrella sampling of particles aggregating on a membrane.

Since SAMBAR is a special case of SATRAM, its performance is briefly discussed for the parallel tempering alanine dipeptide data, in which a comparison to pymbar, a state-of-the-art MBAR solver, is included \cite{shirts2008statistically}.

\subsection*{Using SATRAM to reweight from multiple temperatures}
Alanine dipeptide is simulated in explicit solvent at 21 temperatures ranging from 300K to 500K, spaced 10K apart, see supplementary material for simulation details. 
To apply (SA)TRAM to the parallel tempering dataset, the trajectories are discretized into 40 Markov states using the KMeans++ algorithm \cite{arthur2006k} implemented in the Deeptime software package \cite{hoffmann2021deeptime}. 
To analyze convergence, we compare the respective outputs of (SA)TRAM to a ground truth which we refer to as $\hat{f}_i^k$. This is a converged TRAM estimate computed using Deeptime \cite{hoffmann2021deeptime} to a tolerance of $10^{-10}$.

\begin{figure}
    \centering
    \includegraphics[width=0.85\textwidth]{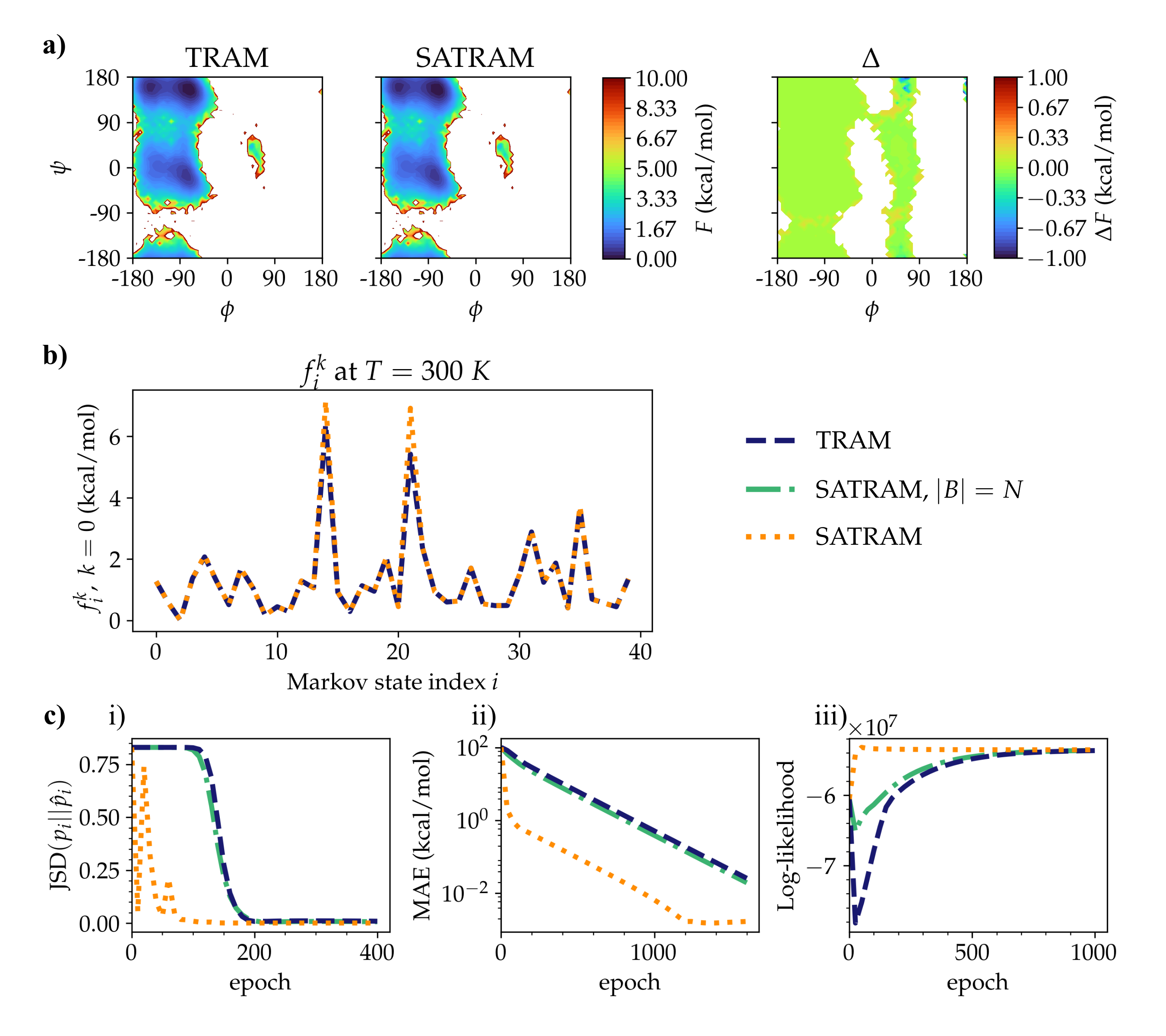}
    \caption{Results for the alanine dipeptide dataset generated by parallel tempering simulations. Data is clustered into 40 Markov states (using KMeans++ clustering). The SATRAM batch size is doubled every $p=10$ epochs.
    \textbf{a)} the free energy profile over the torsion angles estimated by TRAM (left) and SATRAM (center),  and their difference (right),  at 300 K, discretized to a grid of $50\times 50$ bins. Both estimates were run until convergence of the MAE to within 0.1 kcal/mol with respect to $\hat{f}^k$ (1330 epochs for TRAM, 489 for SATRAM).
    \textbf{b)} the estimated $f_i^k$ at $k=0$, i.e. at 300 K. In this state the estimated $f^k_i$ of SATRAM and TRAM show the largest discrepancy.
    \textbf{c)} Convergence of three metrics for TRAM, SATRAM with initial batch size $|B|_0=128$, and SATRAM with $|B|=N$. \textbf{i)} convergence of the JS-divergence of the probability distribution over the Markov states, $p_i$, w.r.t. $\hat{p}_i$. \textbf{ii)}  convergence of the MAE of the $f^k$ w.r.t. $\hat{f}^k$. \textbf{iii)} convergence of the log-likelihood.}
    \label{fig:PT_TRAM}
\end{figure}

The parameters, $f_i^k$ and $v_i^k$, converge towards the ground truth for both SATRAM and TRAM (except for the $i,k$ that correspond to empty states in the case of SATRAM, which are computed in the finalization step). 
\cref{fig:PT_TRAM} (panels a and b) shows the final outputs of (SA)TRAM, after running both methods to within a maximum absolute error (MAE) in the $f^k$ of 0.1 kcal/mol with respect to $\hat{f}^k$, which is well below chemical accuracy. \cref{fig:PT_TRAM}.a show the free energy surface computed at $T=300K$ by TRAM and SATRAM, and the difference between their respective outputs. \cref{fig:PT_TRAM}.b shows the free energies per state, $f_i^k$, for $k=0$, i.e. at 300K. At this temperature, the maximum error between the $f_i^k$ estimated by TRAM and SATRAM is the largest compared to the other temperatures, which is reflected in a slight overestimation of the energy of Markov state at index 14 and 21, two respectively small clusters into which less than 0.2\% of all samples are binned. Not plotted are the $v_i^k$, which are identical for both estimators up to floating-point precision.

The convergence of SATRAM is compared to TRAM, and SATRAM with a batch size $|B|=N$, in \cref{fig:PT_TRAM}.c, which shows the Jensen-Shannon divergence between the probability distributions over the Markov states $p_i=e^{-f_i}$ and $\hat{p}_i = e^{-\hat{f}_i}$ (\cref{fig:PT_TRAM}.c.i), the maximum absolute error between the $f^k$ and $\hat{f}^k$ (\cref{fig:PT_TRAM}.c.ii) and the log-likelihood (\cref{fig:PT_TRAM}.c.iii). For SATRAM the initial batch size was set to $|B|_0=128$, after which the batch size is doubled every $p=10$ epochs, so that $|B|=N$ after 120 epochs, from whereon the implementation is deterministic. From the lines showing SATRAM with $B=|N|$ (in green) we see that the convergence behaviour of the deterministic SATRAM implementation is similar to that of TRAM. 

SATRAM converges faster initially on all metrics in \cref{fig:PT_TRAM}c. The convergence of the MAE shows that SATRAM does not converge to the exact ground truth, which we attribute to accumulating numerical errors:  since the highest energy state is $>6000$ kcal/mol, this is a relative error of the order $10^{-8}$.

The convergence of SATRAM depends primarily on the learning rate/batch size schedule, the number of clusters, and the magnitude free energy difference. A smaller number of clusters will improve the performance of SATRAM over TRAM, and we found increasing the batch size to be more effective than decreasing the learning rate (see supplementary information).

\subsection*{Using SAMBAR to reweight from multiple temperatures}
Since MBAR is a special case of TRAM, it is perhaps unsurprising to see that the convergence behaviour of (SA)MBAR looks similar to that of (SA)TRAM. We compare SAMBAR with an initial batch size $|B|_0=128$ that doubles every $p=10$ epochs to MBAR, and SAMBAR with a batch size $|B|=N$. Additionally, the convergence behaviour of pymbar (using its default adaptive implementation) is compared.

\begin{figure}
    \centering
    \includegraphics[width=0.9\textwidth]{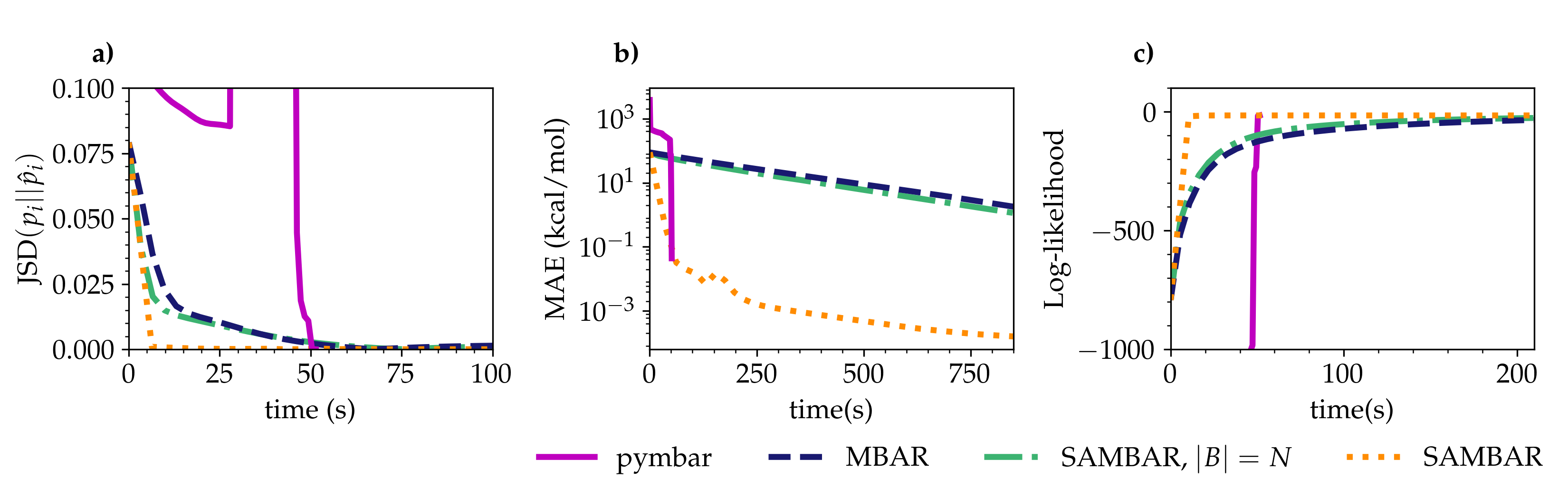}
    \caption{Convergence of three metrics for pymbar, MBAR, SAMBAR with initial batch size $|B|_0=128$, and SAMBAR with $|B|=N$ on the parallel tempering alanine dipeptide dataset. The SATRAM batch size is doubled every $p=10$ epochs.
    \textbf{a)} convergence of the JS-divergence of the probability distribution over 40 bins, $p_i$, w.r.t. $\hat{p}_i$. \textbf{b)} convergence of the MAE of the $f^k$ w.r.t. $\hat{f}^k$. \textbf{c)} convergence of the log-likelihood. }
    \label{fig:PT_MBAR_convergence}
\end{figure}

\cref{fig:PT_MBAR_convergence} shows that SAMBAR converges faster than MBAR on the same three metrics that were used to analyse the convergence of TRAM: the Jensen-Shannon divergence between the probability distributions over 40 clusters (\cref{fig:PT_MBAR_convergence}a) which correspond to the Markov states used for TRAM, the MAE between $f^k$ and $\hat{f}^k$ (\cref{fig:PT_MBAR_convergence}b), and the log-likelihood (\cref{fig:PT_MBAR_convergence}c). The MBAR ground truth, $\hat{f}^k$, is a converged MBAR estimate computed using pymbar \cite{shirts2008statistically}.

\cref{fig:PT_MBAR_convergence} shows that initially, SAMBAR also converges faster than pymbar, although we note that pymbar does not allow for initialization using the bias energies, and the free energies are initialized with zeros. For a fairer comparison of the runtime, the time it takes for pymbar to pre-process the data is not included.  The convergence of SAMBAR may be improved with use of the ADAM \cite{kingma2014adam} optimizer, which results in a performance boost for SAMBAR, allowing it to converge significantly faster than pymbar. A more extensive comparison of SAMBAR, pymbar and FastMBAR is included in the SI.

\subsection*{Using SATRAM to reweight from umbrella sampling}

(SA)TRAM can also be applied to biased simulations, which we demonstrate by applying umbrella sampling to alanine dipeptide. 25 umbrella centers were spaced along both coordinates of interest, the torsion angles $\phi$ and $\psi$ in the protein backbone, giving a total $K=625$ thermodynamic states. For TRAM, the samples are discretized into 5 evenly spaced bins along both torsion angles, so that the number of Markov states $m$ equals $25$.

 An increasing batch size schedule is again employed for SATRAM, starting with an initial batch size $|B|_0 = 128$ and doubling every $p=10$ epochs. The dataset contains 625.000 samples, so that $|B|=N$ after 130 epochs.

\begin{figure}
    \centering
    \includegraphics[width=0.85\textwidth]{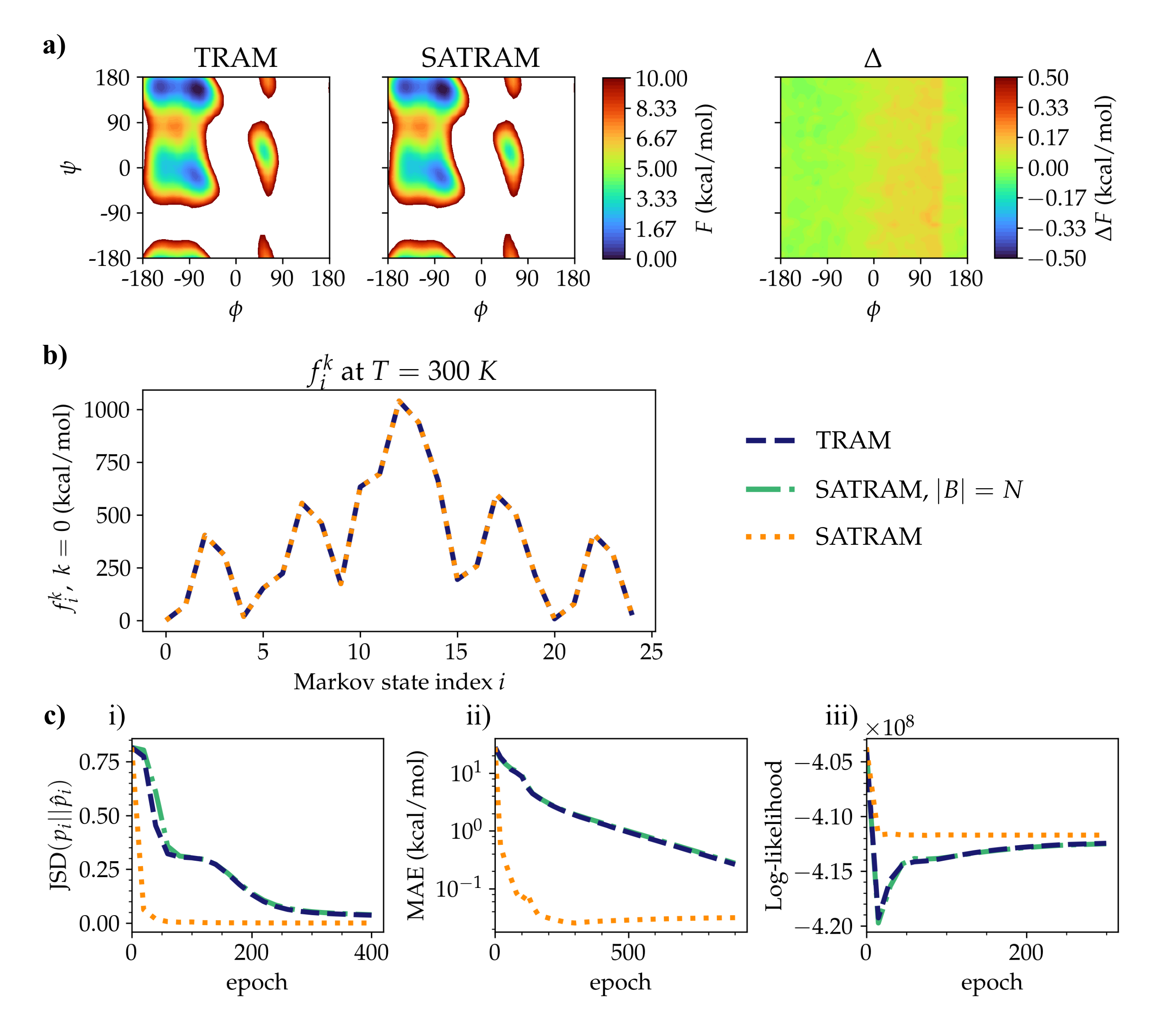}
    \caption{
    Results for the the umbrella sampling dataset of alanine dipeptide with 25 Markov states (using box discretization with 5 bins per torsion angle). The SATRAM batch size is doubled every $p=10$ epochs.
    \textbf{a)} the free energy profile over the torsion angles estimated by TRAM (left) and SATRAM (center),  and their difference (right),  at 300 K, discretized to a grid of $50\times 50$ bins. Both estimates were run until convergence of the MAE to within 0.1 kcal/mol with respect to $\hat{f}^k$ (1179 epochs for TRAM, 79 for SATRAM).
    \textbf{b)} the estimated $f_i^k$ at $k=0$, i.e. at 300 K, in which the graphs of SATRAM and TRAM fully overlap.
    \textbf{c)} Convergence of three metrics for TRAM, SATRAM with initial batch size $|B|_0=128$, and SATRAM with $|B|=N$. \textbf{i)} convergence of the JS-divergence of the probability distribution over the Markov states, $p_i$, w.r.t. $\hat{p}_i$. \textbf{ii)}  convergence of the MAE of the $f^k$ w.r.t. $\hat{f}^k$. \textbf{iii)} convergence of the log-likelihood.
    }
    \label{fig:US_TRAM}
\end{figure}

\cref{fig:US_TRAM} (panels a and b)  shows that the estimates of TRAM and SATRAM converge to the same result for both the estimated free energy surface (\cref{fig:US_TRAM}.a), and the $f_i^k$ (plotted at 300K in \cref{fig:US_TRAM}.b). Not plotted are the $v_i^k$, which are identical for both estimators up to floating-point precision. The Jensen-Shannon divergence (\cref{fig:US_TRAM}.c.i), MAE (\cref{fig:US_TRAM}.c.ii) and log-likelihood (\cref{fig:US_TRAM}.c.iii) converge significantly faster for SATRAM than TRAM.

\subsection*{Using SATRAM to estimate free energy of membrane-mediated interactions}

\begin{figure}
    \centering
    \includegraphics[width=0.78\textwidth]{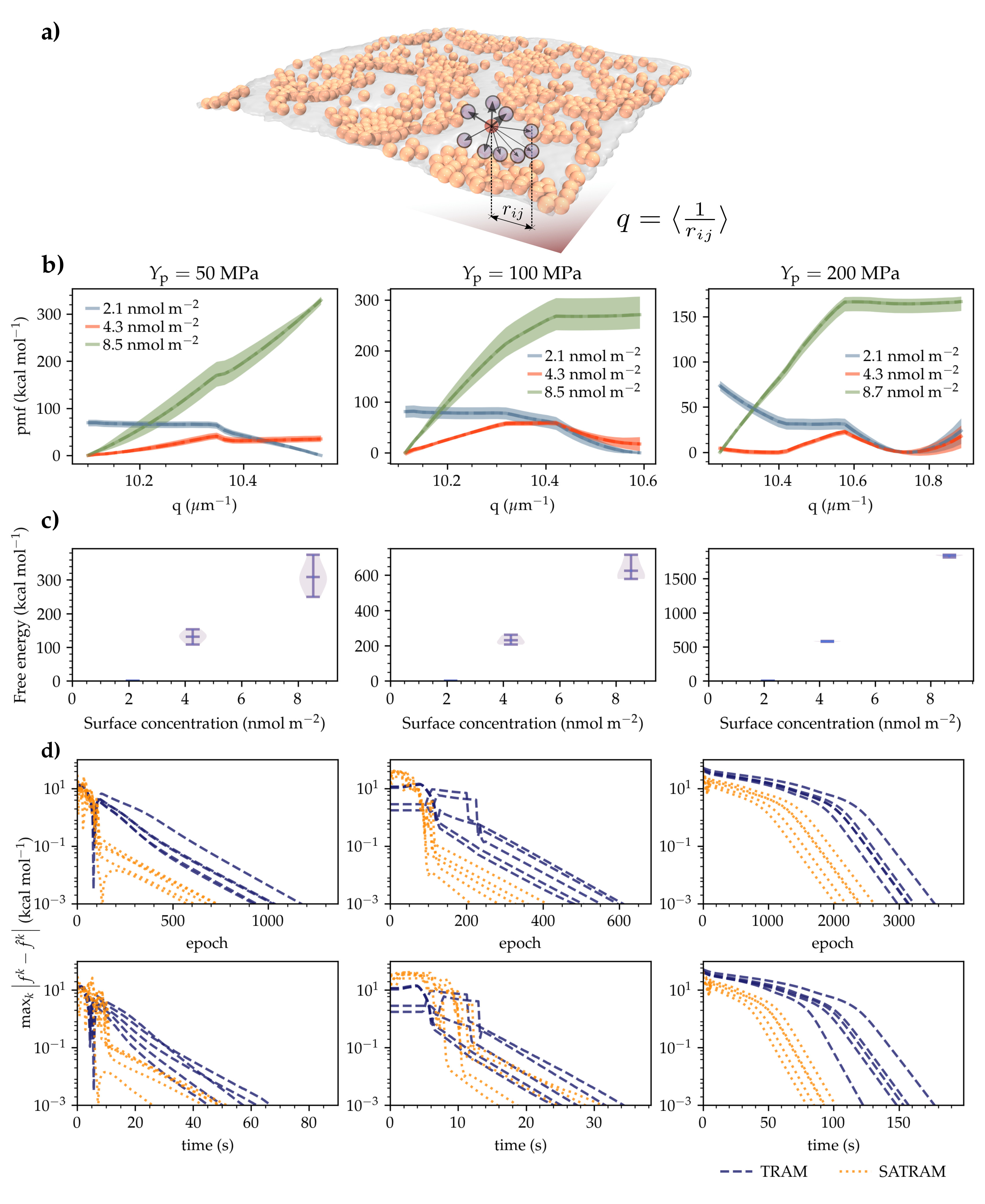}
    \caption{Comparing TRAM and SATRAM in free energy analysis of the aggregation of membrane-bound proteins. \textbf{a)} simulation snapshot with peripheral protein particles shown in orange and membrane particles rendered transparent for clarity. The aggregation reaction coordinate $q$ is defined as the mean of projected pairwise distances between proteins particles. \textbf{b)} potential of the mean force along the reaction coordinate $q$. Results are shown for the given surface concentrations of proteins with three different stiffness values (plot titles). Outputs of the TRAM and SATRAM estimations coincide with the ground truth in the range of inspection (respectively, dashed and dot-dashed lines overlap the solid line). \textbf{c)} comparing free energies of thermodynamic states (i.e. different surface concentrations) between TRAM and SATRAM estimations. The range of free energies highlights the uncertainty in the bias potential. Colors match the legend given at the bottom of the figure. d: Convergence of TRAM and SATRAM with initial batch size $|B|_0=128$, with MAE of the $f^k$ w.r.t. $\hat{f}^k$ used as the metric. Each curve represents an instance of the inferred bias potential.} 
    \label{fig:output_TRAM_SATRAM_MP}
\end{figure}

As a final example, SATRAM is applied to a model of peripheral proteins aggregating on the surface of a lipid bilayer membrane as result of membrane-mediated interactions by \citet{sadeghi2022investigating}. Their work used TRAM to infer the free energy profiles underlying the aggregation/dispersion process. The dataset contains time series at three different surface concentrations (copy number per unit area) $\Gamma_k,\ k=0,1,2,$ of proteins with three different stiffness values $Y_\mathrm{p}$ = 50, 100, 200 MPa. Simulations were performed using a dynamic membrane model \cite{sadeghi2018memmodel,sadeghi2020memhydro} that incorporates flexible peripheral proteins via a consistent force field masking \cite{sadeghi2021thermodynamics}. A reaction coordinate $q$ has been used to quantify aggregation (Fig. \ref{fig:output_TRAM_SATRAM_MP} and supplementary information) with larger $q$ indicating more clustered states. They proposed an effective bias potential of the form $b^k(q) = \xi_k \cdot b(q)$ such that the bias strength $\xi$ correlates with the surface concentration of proteins and $b\left(q\right)$ is a polynomial function specified by protein stiffness. To also estimate uncertainties originating from the assumptions on the bias potential, five independent instances of $\xi_k$ and $b(q)$ functions were derived (details in supplementary information). 

Fig. \ref{fig:output_TRAM_SATRAM_MP} shows SATRAM and TRAM results on free energy estimation in this system. SATRAM and TRAM consistently predict the potential of the mean force along the $q$, coincident with the high-accuracy TRAM estimate computed using Deeptime \cite{hoffmann2021deeptime} to a tolerance of $10^{-10}$ (Fig. \ref{fig:output_TRAM_SATRAM_MP}.b). Mean values of free energies of thermodynamic states, as well as their distributions (due to the uncertainty in the bias potential) match perfectly between SATRAM and TRAM (Fig. \ref{fig:output_TRAM_SATRAM_MP}.c). Convergence curves reveal that SATRAM consistently outperforms TRAM, and can achieve higher accuracy with less computational time. The case is most obvious for the stiffest protein, with the largest difference between free energies of thermodynamic states. This suggests that the stochasticity of SATRAM helps the most with the convergence where the objective function has a more rugged profile due to larger energy gradients.

\section*{Conclusions}

We have introduced SAMBAR and SATRAM, stochastic approximations to MBAR and TRAM respectively, where SAMBAR and MBAR are special cases of the more general SATRAM and TRAM. Like TRAM, SATRAM combines data sampled out of equilibrium in multiple thermodynamic states and estimates a multi-ensemble Markov model, but unlike TRAM, does this in a batch-wise manner. 

By repeatedly doubling the batch size a deterministic implementation is recovered when the batch size reaches the dataset size, allowing for convergence to arbitrary accuracy, but with an initial performance boost. We showed that SATRAM converges to chemical accuracy up to an order of magnitude faster than TRAM. 
The batch-wise formulation also allows for adding data on the fly, so that the free energies may be computed in parallel with the sampling algorithm.

The SATRAM code is available on github at \href{https://github.com/noegroup/SATRAM}{https://github.com/noegroup/SATRAM}.

\begin{acknowledgement}
We thank Moritz Hoffmann, Phillip S. Hudson, Tim Hempel, and Yaoyi Chen for offering test data and helpful discussions.
We gratefully acknowledge funding from European Commission (Grant No. ERC CoG 772230 “ScaleCell”), the BMBF (Berlin Institute for Learning and Data, BIFOLD), the Berlin Mathematics center MATH+ (AA1-6), the Deutsche Forschungsgemeinschaft DFG (SFB1114/A04,C03 and SFB958/A04), the NSF of China (Grant No. 12171367) and the Shanghai Municipal Science and Technology Commission (Grant No. 20JC1413500, 21JC1403700 and 2021SHZDZX0100).

\end{acknowledgement}

\nocite{eastman2017openmm,bgmol}
\bibliography{bibliography}

\setcounter{section}{0}
\setcounter{equation}{0}
\setcounter{figure}{0}

\renewcommand{\thesection}{S.\arabic{section}}
\renewcommand{\thefigure}{S.\arabic{figure}}
\renewcommand{\theequation}{S.\arabic{equation}}

\begin{suppinfo}
Listed in the supporting information are the derivation of SAMBAR and SATRAM, implementation details, details on the simulation setup for generating the data used. Additionally, in the SI, the dependence of SATRAM/SAMBAR on various hyperparameters is examined in detail, and SAMBAR is compared to selected existing MBAR implementations.

\section{SAMBAR derivation}
\cref{eq:MBAR_self_consistent} can equivalently be written as
\begin{equation}
\frac{1}{N} \sum_{x\in X} \frac{N \; \text{exp}[f^k-b^k(x)]}{\sum_{l=1}^S N^l \;\text{exp}[f^l - b^l(x)]} = 1,
\end{equation}
so that for one random sample in our dataset, we have the expectation value
\begin{equation}
\mathbb{E}  \left[\frac{N \; \text{exp}[f^k-b^k(x)]}{\sum_{l=1}^S N^l \;\text{exp}[f^l - b^l(x)]} - 1 \right] = 0. \label{expectation_f_MBAR}
\end{equation}
This allows the use of stochastic approximation of $f^k$ \cite{robbins1951stochastic}. The update for a stochastic iteration using a batch of samples $B$ is thus given by
\begin{equation}
f^k = f^k - \eta \left( \frac{1}{|B|} \sum_{x\in B} \frac{N \; \text{exp}[f^k -b^k(x)]}{\sum_{l=1}^S N^l \;\text{exp}[f^l - b^l(x)]}\right),
\end{equation}
where $\eta$ is the learning rate. The additive term has been dropped since we only want to determine the free energies up to an additive constant. To avoid a large drift in the energies, a normalization step is executed after each free energy update, which consists of shifting all $f^k$ by a constant so that the minimum of the
free energies equals zero.

\section{SATRAM derivation}
Using \cref{eq:TRAM_1,eq:TRAM_2,eq:reduced_sample_counts}, let $v_i^k := v_i^k /N$, $R_i^k := R_i^k/N$, to obtain the respective equivalent equations
\begin{align}
\frac{1}{N} \sum_j \frac{(c_{ij}^k + c_{ji}^k) v_i^k}{\mathrm{exp}[f_j^k - f_i^k] v_j^k + v_i^k} = v_i^k, \quad \text{for all } \;i,\; k, \\
\frac{1}{N} \sum_{x \in X} 1_{x \in S_i} \frac{\mathrm{exp}[f_i^k -b^k(x)]}{\sum_l R_i^l\; \mathrm{exp}[f_i^l - b^l(x)]} = 1, \quad \text{for all }\; i, \;k, \label{TRAM_2_2}\\
R_i^k = \frac{1}{N} \left( \sum_j \frac{(c_{ij}^k + c_{ji}^k)v_j^k}{v_j^k+\mathrm{exp}[f_i^k -f_j^k]v_i^k} + N_i^k - \sum_j c_{ji}^k \right). \label{TRAM_2_3}
\end{align}
Drawing one random sample $x$ from our dataset, using \cref{TRAM_2_2}, we can write the expectation value
\begin{equation}
    \mathbb{E} \left[ 1_{x \in S_i}  \frac{\mathrm{exp}[f_i^k -b^k(x)]}{\sum_l R_i^l\; \mathrm{exp}[f_i^l - b^l(x)]} \right] = 1. \label{eq:expectation_f_TRAM}
\end{equation}
We use \cref{eq:expectation_f_TRAM} to approximate $f_i^k$ by stochastic approximation \cite{tan2017optimally}, where we can again drop the additive term since we want to compute the free energies up to an additive constant. The $f_i^k$ and $v_i^k$ are updated iteratively as  
\begin{align}
    f_i^k &:= f_i^k - \eta \left(\frac{1}{|B|} \sum_{x \in B} 1_{x\in S_i}  \frac{\mathrm{exp}[f_i^k-b^k(x)]}{\sum_l R_i^l\; \mathrm{exp}[f_i^l - b^l(x)]} \right), \label{eq:SATRAM_A1}\\
    v_i^k &:= (1-\eta)\cdot v_i^k + \eta  \cdot \frac{1}{N} \sum_j \frac{(c_{ij}^k + c_{ji}^k)v_i^k}{\mathrm{exp}[f_j^k - f_i^k] v_j^k + v_i^k},
\end{align}
where the $R_i^k$ are computed as \cref{TRAM_2_3}. 
As with SAMBAR, a normalization step is performed after each iteration to avoid a drift in the energies.
We note that when setting $\eta = 1$ and $B=X$, the optimal $f_i^k$ and $v_i^k$ form a fixed point of iteration.

\section{Implementation notes}
For TRAM all calculations are performed in log-space to avoid over-and underflows of the exponentials. The free energy updates for SATRAM cannot be computed in log-space, which can result in a numerical underflow when there are large differences between the free energies of the different states (of the order of $1000$ kcal/mol), which can occur when one runs SATRAM from an initial estimate of $f_i^k=0\; \forall f, i$. A simple solution for this problem is starting estimation from a `well-educated guess', namely the mean of the bias energies, i.e. $f_i^k = \langle b^k(x) \rangle, \; \forall k, i$. This also significantly reduces the running time of the estimation.

Cutting off the absolute values of the free energy update for SATRAM improves convergence. $\Delta f_i^k = \mathrm{min}(|\eta \cdot\Delta f_i^k|,\; r)$ where $r$ is the cutoff value, and $\Delta f_i^k$ is the expression between the brackets of \cref{eq:SATRAM_1}. Decreasing the learning rate over time will then have the added effect of increasing the effective cutoff value.

\section{Simulation details}

\subsection*{Alanine dipeptide}
All data were generated using OpenMM (version 7.5.1) \cite{eastman2017openmm}. The system's initial conditions were a calibrated system of alanine dipeptide solvated in water that can be downloaded from the bgmol repository on GitHub \cite{bgmol}. 

\paragraph{Parallel tempering}
The parallel tempering simulations consist of 21 temperatures spaced 10K apart over the interval (300K, 500K).

Each system is integrated in an NVT ensemble using a Nose Hoover integrator with step size $2.0\;\mathrm{fs}$ and friction $1.0\;\mathrm{ps}^{-1}$. Before production, local energy is minimized and the system equilibrated over a $100\;\mathrm{ps}$ timespan. Samples are taken in $1\;\mathrm{ps}$ intervals. For the lowest six temperatures (300-350K, inclusive) 50000 samples are taken per state, and for the higher temperatures (350-500K, inclusive) 10000 samples per state, to a total of 450.000 samples.

\paragraph{Umbrella sampling}
The bias potentials are of the form $U(\psi, \phi) = \frac{1}{2}K [(\psi-\psi_0)^2 + (\phi-\phi_0)^2]$, where $K=200 \;\mathrm{kJ/mol}\cdot \mathrm{(rad)}^2$, and with $\theta_0$ evenly distributed in interval $(-\pi, \pi)$. For both $\phi$ and $\psi$, 25 bias centres are used, resulting in a total of 625 thermodynamic states.

Each system is integrated in an NVT ensemble, using a Langevin integrator at $T=300\;K$ with step size $2.0\;\mathrm{fs}$ and friction coefficient $1.0\;\mathrm{ps}^{-1}$. After applying bias potentials, local energy is minimized and the system is equilibrated over 200 ps. Samples are taken in $1 \; \mathrm{ps}$ intervals. 1000 samples are taken per thermodynamic state, to a total of $625.000$ samples.

\subsection*{Peripheral membrane proteins simulation details}
Simulation data on the aggregation of membrane peripheral proteins are the same as used in the study of entropic membrane-mediated interactions by \citealt{sadeghi2022investigating}. These simulation were performed using a highly coarse-grained membrane model that represents the two membrane leaflets via particle dimers and relies on bonded interactions between nearest neighbor particles to mimic the mechanics of a bilayer membrane \cite{sadeghi2018memmodel}. In-plane fluidity is implemented via bond-flipping Monte Carlo moves, that are calibrated to reproduce the observed lateral dynamics of proteins such as Shiga and Cholera toxins bound to the membrane \cite{sadeghi2021thermodynamics}. The model additionally includes realistic out-of-plane kinetics of a membrane suspended in aquatic solvent via a hydrodynamic coupling method developed in tandem \cite{sadeghi2020memhydro}. Peripheral proteins are added to the model as particles tagged to introduce a locally masked force field. The force field parametrization reflects an intrinsic curvature as well as a contributing stiffness for each protein. 

All simulations were performed at $T$ = 310 K using an anisotropic over-damped Langevin integrator with hydrodynamic interactions, with a time step of 0.15 ns. An in-plane stochastic barostat was used to keep membrane patches tensionless. The dataset contains a total of 9 trajectories, pertaining to three surfaces concentrations of proteins with three different stiffness values. In each simulation, the system is equilibrated for 1 ms and samples were taken for another 2 ms at 0.3 $\mu$s intervals. All simulations were performed using a specific-purpose software developed based on the membrane model.

\paragraph{Aggregation reaction coordinate}
The reaction coordinate $q$ used to describe the degree of aggregation on the surface of the membrane is defined based on the inverse pairwise distances between proteins \cite{sadeghi2021thermodynamics, sadeghi2022investigating},
\begin{equation}
    \label{eq:reaction_coord}
        q\left(t\right) = \frac{2}{N_\mathrm{p}\left(N_\mathrm{p} - 1\right)}\sum_{i, j>i}\frac{1}{r_{ij} \left(t\right)}
\end{equation}
where $N_\mathrm{p}$ is the number of proteins bound to the membrane, and $r_{ij}$'s are the pairwise distances between particle positions projected to a 2D plane parallel to the initial (undeformed) mid-surface of the membrane.

\paragraph{Bias potential}
The application of the masked force field only modifies the interaction potential without changing the number of particles. Thus, the assumption was made that the Hamiltonian is biased as $\mathcal{H} = \mathcal{H}_0 + \xi\, b\left(q\right)$, with $\xi$ proportional to the surface concentration. Justification for this assumption on the bias potential was previously demonstrated \cite{sadeghi2021thermodynamics}, with Weighted Histogram Analysis Method (WHAM) used to obtain free energy landscapes consistent with the expected nucleation dynamics \cite{sadeghi2021thermodynamics}. 

For the simulations presented here, polynomials of degree 5 are used as the function $b\left(q\right)$. A total of 5 different bias functions were obtained using least-square fits to randomly sub-sampled chunks of energy data. These ``instances'' of bias potential have been independently used to estimate free energy profiles, thus yielding the presented uncertainties in free energies.

\section{Optimizing SATRAM}
The effect of reducing the learning rate over time or increasing the batch size on the convergence rate is investigated. We find that increasing the batch size over time is much more effective than decreasing the learning rate. An initial learning rate of $\eta_0 = \sqrt{\frac{|B|}{N}}$ works well for all datasets.

\subsection*{Decreasing the learning rate}

\cref{fig:PT_TRAM_decrease_lr} shows convergence profiles for SATRAM with different learning rate schedules. For all SATRAM estimates, the batch size was set to 8192 and the initial learning rate to $\eta_0 = \sqrt{\frac{|B|}{N}}$. The learning rate was decreased by a factor $\gamma$ every $p$ epochs. From the figures we can see that decreasing the learning rate is not effective, as a larger learning rate leads to a biased estimate of the free energies, and a smaller learning rate reduces the convergence rate to such an extent that SATRAM converges much slower than TRAM. The standard error of the MAE is of the order $10^{-3}$ kcal/mol. Error bars would not be visible in most regions of the plot and are therefore omitted.

\begin{figure}
    \centering
    \includegraphics[width=.85\textwidth]{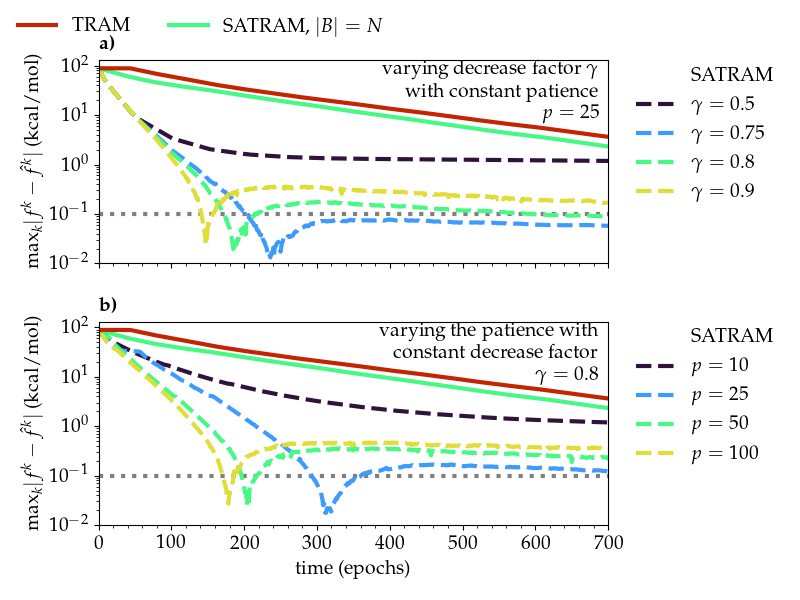}
    \caption{The convergence of the MAE for TRAM and SATRAM on the parallel tempering dataset with $K=40$ clusters. Top: the learning rate is decreased by a factor $\gamma$ every $p=10$ epochs. Bottom: learning rate is decreased by a factor $\gamma=0.8$ every $p$ epochs. Batch size was set to $|B|=8192$ for all estimates. The grey line represents chemical accuracy. The finalization step of SATRAM was not taken into account and could decrease SATRAM errors by a small margin after estimation.}
    \label{fig:PT_TRAM_decrease_lr}
\end{figure}

The distinct shape of the convergence arises due to the fact that for this dataset, the SATRAM estimate first crosses the true value of the free energies after which it converges from below, whereas TRAM approaches it from above (see \cref{fig:PT_TRAM_convergence_shape_cause}).

\begin{figure}
    \centering
    \includegraphics[width=0.75\textwidth]{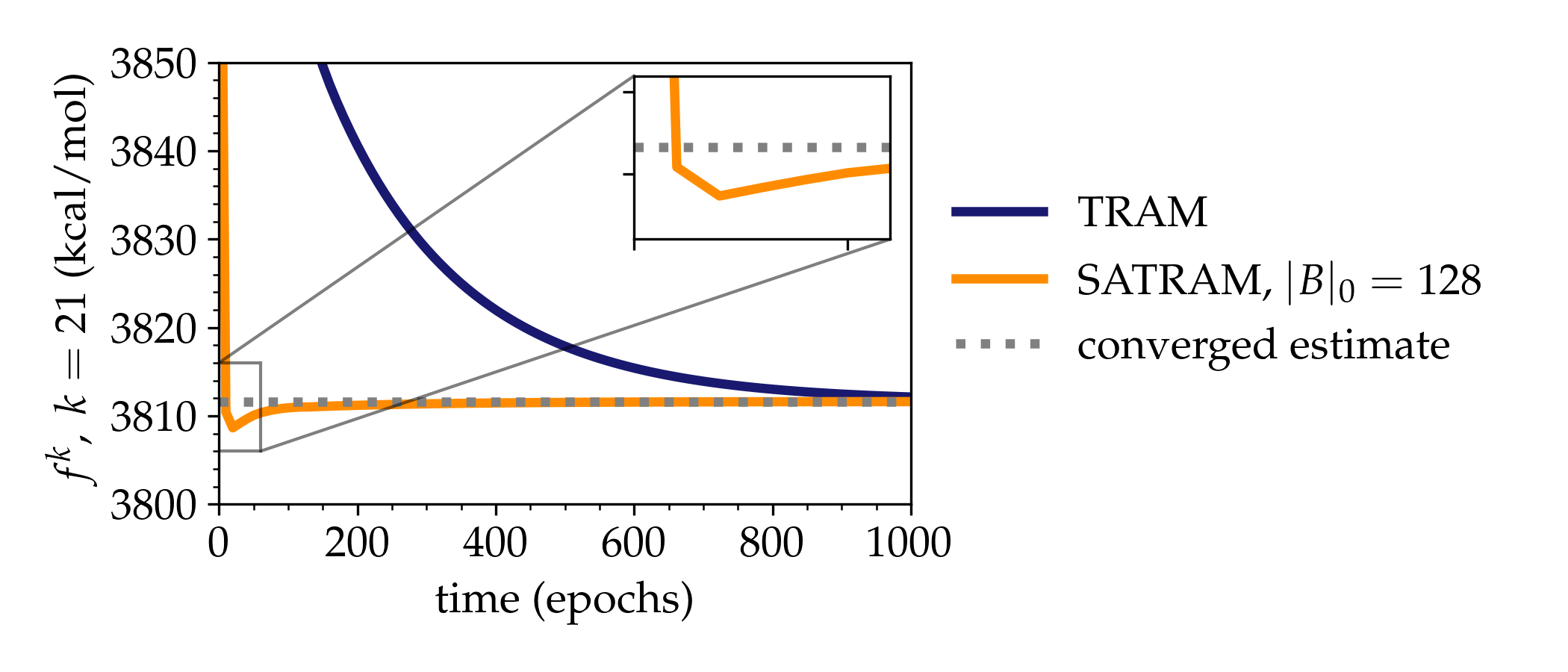}
    \caption{The convergence of the highest free energy state, for TRAM, SATRAM with batch size set to 8192, on the parallel tempering dataset with $K=40$ clusters. The SATRAM estimate `crosses' the true value of the free energy, producing the spikes in the convergence of SATRAM as seen in  \cref{fig:PT_TRAM_decrease_lr,fig:PT_TRAM_convergence_batch_size,fig:PT_TRAM_convergence_batch_size_increase}
    }
    \label{fig:PT_TRAM_convergence_shape_cause}
\end{figure}

\subsection*{Increasing the batch size}
\begin{figure}
    \centering
    \includegraphics[width=0.85\textwidth]{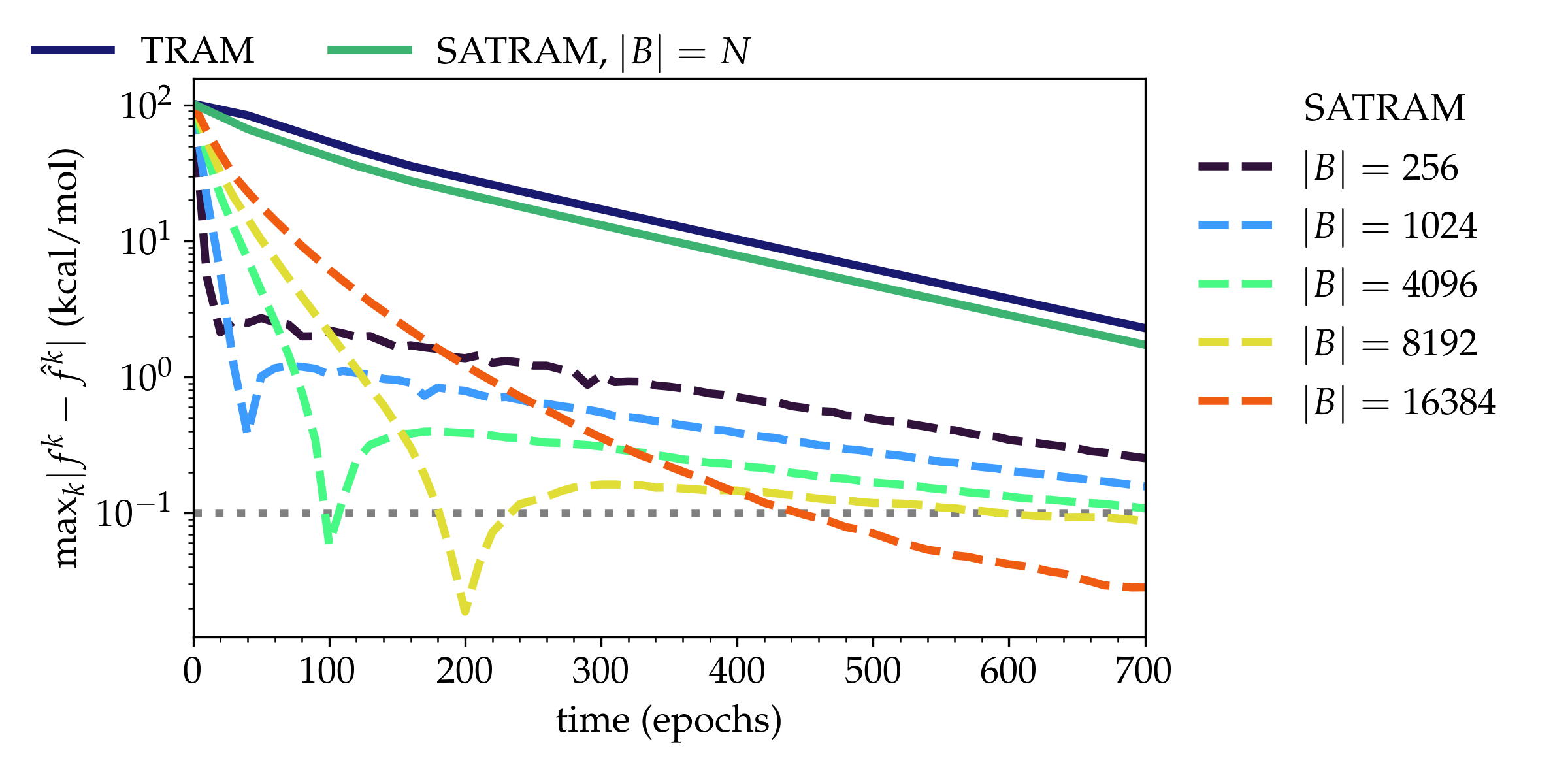}
    \caption{The convergence of the MAE between the estimated $f^k$ and $\hat{f}^k$, for TRAM, and SATRAM with various batch sizes, on the parallel tempering dataset with $K=40$ clusters. The initial learning rate was set to $\eta_0 = \sqrt{\frac{|B|}{N}}$ and decreased by a factor $\gamma=0.9$ every $25$ epochs. The grey line represents chemical accuracy. The finalization step of SATRAM was not taken into account and could decrease SATRAM errors by a small margin after estimation.}
    \label{fig:PT_TRAM_convergence_batch_size}
\end{figure}

 \cref{fig:PT_TRAM_convergence_batch_size} shows convergence rates for various batch sizes in combination with a decreasing learning rate schedule. From \cref{fig:PT_TRAM_convergence_batch_size} it becomes apparent that there is a benefit to increasing the batch size over time, as a smaller batch size leads to a faster decrease in the error initially, whereas a larger batch size converges faster at a later stage of estimation.
\begin{figure}
    \centering
    \includegraphics[width=0.85\textwidth]{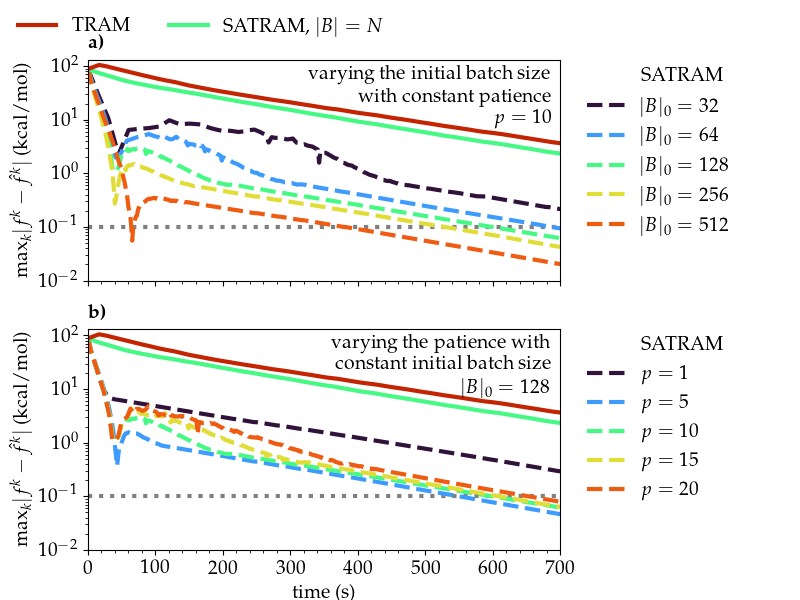}
    \caption{The convergence of the MAE for TRAM and SATRAM on the parallel tempering dataset with $K=40$ clusters. Top: the batch size is doubled every $p=10$ epochs, with a varying initial batch size $|B|_0$. Bottom: the initial batch size is set to $|B|_0=128$ and the batch size is doubled every $p$ epochs. The grey line represents chemical accuracy. The finalization step of SATRAM was not taken into account and could decrease SATRAM errors by a small margin after estimation.}
    \label{fig:PT_TRAM_convergence_batch_size_increase}
\end{figure}

\cref{fig:PT_TRAM_convergence_batch_size_increase} shows the convergence of different schedules where the batch size is doubled every $p$ epochs, until it reaches $|B|\geq N$, after which the implementation becomes deterministic. The learning rate is re-computed to $\eta=\sqrt{\frac{|B|}{N}}$ after the batch size is increased. Additionally decreasing the learning rate by a factor $\gamma$ was found not to improve convergence.

\subsection*{Dependence on the number of clusters}

\begin{figure}
    \centering
    \includegraphics[width=\textwidth]{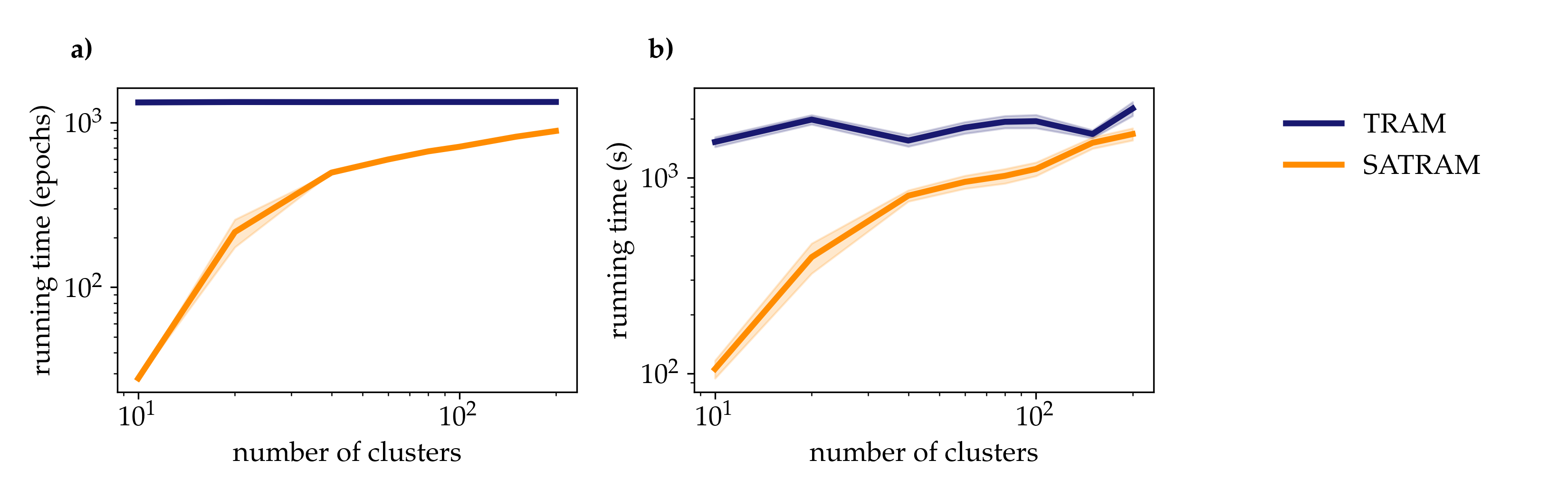}
    \caption{The time it takes in epochs (left) and seconds (rigth) for the MAE to converge to $\leq 0.1$ kcal/mol for TRAM and SATRAM on the parallel tempering dataset, depending on the number of clusters $K$. The line and shaded area represent the mean and standard error of 10 estimates with different random seeds respectively. SATRAM was run with an increasing batch size starting at $|B|_0=128$ and doubling every $p=5$ epochs. The finalization step of SATRAM was not taken into account for convergence.}
    \label{fig:PT_TRAM_convergence_cluster_dependence}
\end{figure}

The convergence of TRAM does not depend on the number of cluster centers but rather on the largest free energy difference, which is fairly constant with respect to the number of cluster centers. For SATRAM however, a larger number of cluster centers leads to relatively poorer performance since there will be a larger variance in the free energy updates due to the larger number of parameters (see \cref{fig:PT_TRAM_convergence_cluster_dependence}). We recommend setting the initial batch size to $|B|_0 \geq K$.

\section{Optimizing SAMBAR}
Since the SAMBAR equations do not depend on Lagrangian multipliers $v_i^k$, batch-wise optimization can easily be accelerated by optimizers that include momentum, such as ADAM \cite{kingma2014adam}. This will improve the rate of convergence to such a rate that SAMBAR outperforms pymbar \cite{shirts2008statistically} and is on par with FastMBAR \cite{ding2019fast}, both of which are implemented using second-order optimization algorithms (see \cref{fig:US_MBAR_convergence}).
For this figure, SAMBAR was implemented using an Adam optimizer in combination with two different learning rate schedulers, of which the scheduling depended on the batch size. Given the batch size $|B|$, the initial learning rate was set to $\eta_0=\sqrt{\frac{|B|}{/N}}$. Given the number of batches per epoch $\Gamma=\frac{N}{|B|}$, the first scheduler decreased the learning rate on each plateau lasting $\Gamma/2$ batches by a factor 0.5, down to a learning rate of $\eta=0.1\eta_0$. The second scheduler decreased the learning rate by a factor 0.9 on each plateau lasting $\Gamma$ batches.
\begin{figure}
    \centering
    \includegraphics[width=\textwidth]{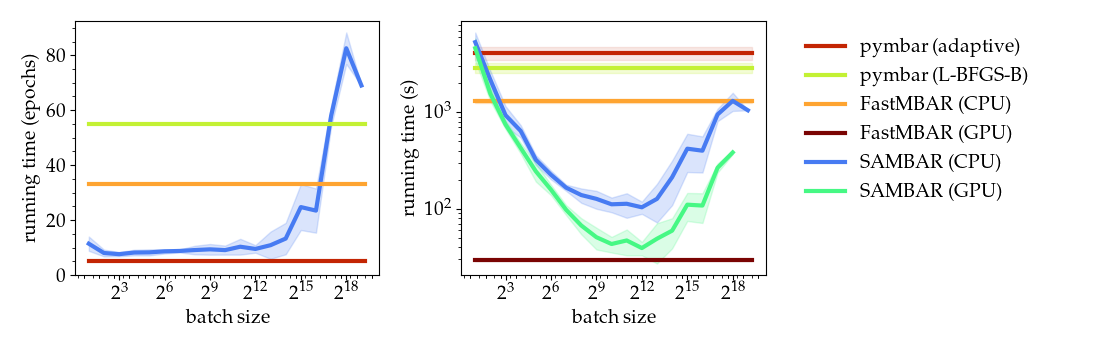}
    \caption{Running time comparison in epochs (left) and seconds (right) of FastMBAR (L-BFGS-B implementation on CPU and GPU), pymbar (adaptive and L-BFGS-B implementations on CPU), and SAMBAR. The SAMBAR running time depends on the chosen batch size, $|B|$, whereas the other determinations are deterministic (plotted as horizontal lines). 
    \\
    Measures is the time it takes for the MAE between the estimated $f^k$ and $\hat{f}^k$ is $\leq0.1$ kcal/mol, on the umbrella sampling dataset. The left graph does not differentiate between CPU and GPU implementations since the number of epochs is the same on both. The lines and shaded areas represent the mean and standard error of 10 estimates with different random seeds respectively. }
    \label{fig:US_MBAR_convergence}
\end{figure}
\end{suppinfo}

\end{document}